\documentclass[aps,preprint,showpacs,showkeys,nofootinbib,floatfix,superscriptaddress]{revtex4}
\usepackage{epsfig}
\usepackage{graphicx}
\usepackage{multirow}
\usepackage{amsmath,stackrel}
\usepackage{bm}
\usepackage{soul}
\usepackage{color}
\newcommand{\be}{\begin{equation}}
\newcommand{\ee}{\end{equation}}
\newcommand{\beq}{\begin{equation}}
\newcommand{\eeq}{\end{equation}}
\newcommand{\bea}{\begin{eqnarray}}
\newcommand{\eea}{\end{eqnarray}}
\newcommand{\ave}[1]{\langle {#1} \rangle}
\newcommand{\tave}[1]{\langle\!\langle{#1}\rangle\!\rangle}
\begin{document}
\title{Charmed baryons in nuclear matter}

\author{T.~F.~Caram\'es}
\email{carames@usal.es}
\affiliation{Departamento de F\'\i sica Fundamental e IUFFyM, Universidad de Salamanca, E-37008
Salamanca, Spain}
\author{C.~E.~Fontoura}
\email{ce.fontoura@unesp.br}
\affiliation{Instituto Tecnol\'ogico de Aeron\'autica, DCTA, 12228-900 S\~ao Jos\'e dos Campos,
SP, Brazil}
\author{G.~Krein}
\email{gastao.krein@unesp.br}
\affiliation{Instituto de F\'{\i}sica Te\'{o}rica, Universidade Estadual
Paulista, Rua Dr. Bento Teobaldo Ferraz, 271 - Bloco II, 01140-070 S\~ao Paulo, SP, Brazil}
\author{J.~Vijande}
\email{javier.vijande@uv.es}
\affiliation{Unidad Mixta de Investigaci\'on en Radiof{\'\i}sica e Instrumentaci\'on Nuclear 
en Medicina (IRIMED), Instituto de Investigaci{\'o}n Sanitaria La Fe (IIS-La Fe)-Universitat de Valencia (UV) 
and IFIC (UV-CSIC), Valencia, Spain}
\author{A.~Valcarce}
\email{valcarce@usal.es}
\affiliation{Departamento de F\'\i sica Fundamental e IUFFyM, Universidad de Salamanca, E-37008
Salamanca, Spain}

\begin{abstract}
We study the temperature and baryon density dependence of the masses of the lightest charmed  
baryons $\Lambda_c$, $\Sigma_c$ and $\Sigma^*_c$. We also look at the effects of the
temperature and baryon density on the binding energies of the $\Lambda_c N$ and $\Lambda_c \Lambda_c$ 
systems. Baryon masses and baryon-baryon interactions are evaluated within
a chiral constituent quark model. Medium effects are incorporated in those parameters of the model
related to the dynamical breaking of chiral symmetry, which are the masses of the constituent quarks, 
the $\sigma$ and $\pi$ meson masses, and quark-meson couplings. We find that while the in-medium 
$\Lambda_c$ mass decreases monotonically with temperature, those of $\Sigma_c$ and $\Sigma^*_c$ have 
a nonmonotonic dependence. These features can be understood in terms of a 
simple group theory analysis regarding the one-gluon exchange interaction in those hadrons. The in-medium 
$\Lambda_c N$ and $\Lambda_c \Lambda_c$ interactions are governed by a delicate balance involving 
a stronger attraction due to the decrease of the $\sigma$ meson mass, suppression of coupled-channel 
effects and lower thresholds, leading to shallow bound states with binding energies of a few~MeV. 
The $\Lambda_c$ baryon could possibly be bound to a large nucleus, in qualitative agreement with results 
based on relativistic mean field models or QCD sum rules. Ongoing experiments at RHIC or LHCb
or the planned ones at FAIR and J-PARC may take advantage of the present results. 
\end{abstract}


\maketitle

%
\section{Introduction}
\label{secI}						
The study of hadrons containing charm quarks is of broad interest nowadays in our quest to
understand the fundamental theory of the strong interaction, quantum chromodynamics (QCD). 
Spectroscopy of such hadrons, which has been at the forefront of research since the discovery of the charm 
quark in~1974, continues receiving most of the attention in light of the continuous discovery 
during the last decade of the so--called $XYZ$ exotic mesons, most of which have been observed above 
the charm production threshold. Their internal structure is still under scrutiny, the main
conjectures being threshold effects, molecular structures or compact multiquark 
states~\cite{{Esposito:2016},Ric16,{Briceno:2015rlt},{Chen:2016},{Lebed:2016hpi},Ali17,{Guo:2017}}. The major
difficulty here, as in any other instance involving hadron spectroscopy, is our poor understanding of 
the low energy regime of QCD, which is dominated by nonperturbative phenomena like color confinement and 
mass generation. In this respect, the study of charmed hadrons in matter offers 
an extremely promising possibility for exploring such phenomena. In a medium composed predominantly by light quarks, like 
an atomic nucleus whose properties are determined by the nonperturbative physics at the energy scale 
$\Lambda_{\rm QCD}$, the charm quark plays the role of an {\em impurity particle} because of its large mass, 
$m_c \simeq 5~\Lambda_{\rm QCD}$~\cite{Pat16}. This is to say that the vacuum properties of a charm 
quark are little or not at all modified in an atomic nucleus and any change in the properties of a
hadron containing charm can be linked to its light constituents. Indeed, the proposal made long time 
ago of using charmonia as probes of the properties of the excited matter produced 
in a relativistic heavy ion collision is a prime example of such a possibility~\cite{Matsui:1986dk}. Other possibilities 
include $D-$mesic nuclei, charmed hypernuclei and nuclear-bound charmonia~\cite{{Hosaka:2016ypm},
{Krein:2017usp}}.  

In recent years, there has been an impressive experimental progress in the spectroscopy of heavy 
baryons, mainly in the charm sector. The LHCb Collaboration at the Large Hadron Collider (LHC),
in particular, is engaged in an extensive program aimed at analyses of charmed hadrons produced 
in the environment of high-energy proton-proton collisions~\cite{Ogi15} and has already 
reported the observation of five new narrow excited $\Omega_c$ states~\cite{Aaij:2017nav}. Also, 
coalescence and statistical hadronization models~\cite{Cho:2017dcy} predict that not only charmed 
baryons $Y_c = (\Lambda_c, \Sigma_c, \Xi_c, \Omega_c, \cdots)$, but also $Y_c N$ bound or resonant 
states, where $N=(p,n)$, are produced at relatively high rates in the environment 
of a heavy-ion collision at the LHC. In addition, in the coming years 
experiments aimed at producing charmed hypernuclei, in which a $Y_c$ baryon is bound to a nucleus, 
are becoming realistic at the planned installation of a 50~GeV high-intensity proton beam at Japan 
Proton Accelerator Research Complex (J-PARC)~\cite{Nou17,Fuj17}. 
There are also planned experiments by the $\overline{\rm P}$ANDA Collaboration at the Facility for 
Antiproton Ion Research (FAIR)~\cite{Wie11,Hoh11} to produce charmed hadrons by annihilating 
antiprotons on nuclei. 

These experimental prospects have reinvigorated studies of the low-energy $Y_c N$ interactions in free
space and also in matter~\cite{Oka13,{Tsushima:2002ua},{Cai03},{Tsu04},{Tan04},{Kop07},{Wan11},{Wan12},
{Liu12},{Hua13},{Gar15},{Mae16},{Fontoura:2017gpu},{Haidenbauer:2017dua},{Miyamoto:2017tjs},
{Miyamoto:2017ynx},{Azi17},{Azizi:2018dtb},{Ohtani:2017wdc},{Chen:2017vai},{Meng:2017udf},{Maeda:2018xcl}}. 
More recently, also $Y_c Y_c$ interactions are becoming focus of 
interest~\cite{Lee11,Car15,Hua14,{Meguro:2011nr},{Li:2012bt},{Zhao:2013ffn},{Meng:2017fwb},{Chen:2018pzd},{Yang:2018amd}}. 
The complete lack of experimental information on the elementary $Y_c N$  and $Y_c Y_c$ interactions in free space imposes 
great difficulties in accessing in-medium effects. 

Although recent results from lattice simulations of the $Y_c N$ interactions provide 
important guidance~\cite{{Miyamoto:2017tjs},{Miyamoto:2017ynx}}, they are still obtained with unphysical pion masses 
and need to be extrapolated using, e.g., a chiral effective field theory~\cite{Haidenbauer:2017dua}. 
In order to make progress, the strategies used so far rely on the use of models constrained as much as 
possible by symmetry arguments, analogies with similar systems, and the use of different degrees of 
freedom. 

Relativistic mean field models, based either on quark degrees of freedom~\cite{Saito:2005rv}
or hadronic degrees of freedom~\cite{Hagino:2014zua} have been widely used to study medium dependence 
of the strange hyperons $Y = (\Lambda, \Sigma, \cdots)$ and are natural candidates to be extended to 
the charm sector. Calculations using quark degrees of freedom~\cite{Tsu04,Tan04} predict that some 
of the $Y_c$ baryons are likely to be bound to sufficiently large nuclei. A similar 
calculation~\cite{Fontoura:2017gpu}, but based on a nonlinear meson-exchange mean field model, 
predicts that $\Lambda_c$ can be bound to nuclei as small as $^{12}{\rm C}$. 

QCD sum rules is 
another technique that has been used to investigate properties of charmed baryons in nuclear matter, 
sometimes with contradictory conclusions. One major source of uncertainty with
the QCD sum rules is the lack of reliable information on the in-medium condensates, i.e., the
thermodynamical average of products of quark and gluon operators. Although the lowest-dimension
in-medium condensates are relatively well constrained by chiral symmetry and the trace anomaly, 
condensates of higher dimension are commonly treated in the so-called factorization hypothesis, in that
they are written as products of the lowest-dimension condensates~\cite{Cohen:1994wm}. For example, 
using the factorization hypothesis Ref.~\cite{Wan11} finds that the mass of the $\Lambda_c$ increases 
in nuclear matter, which means that $\Lambda_c$ feels an average repulsive potential. The same author concludes 
in Ref.~\cite{Wan12} that the mass of the $\Sigma_c$ baryons decreases, while Ref.~\cite{Azizi:2018dtb} 
finds the opposite result; both calculations use the factorization hypothesis. Still, Ref.~\cite{Azi17}, 
also using the factorization hypothesis, finds that the mass of $\Lambda_c$ decreases considerably in nuclear 
matter. Ref.~\cite{Ohtani:2017wdc} examines critically the role played by the factorization hypothesis and finds,
in particular, that the results based on QCD sum rules depend strongly on 
the density dependence of the four-quark condensate. When using a density dependence based on a 
factorization hypothesis for the four-quark condensate, that reference 
predicts that the $\Lambda_c$ mass increases with density, while the opposite behavior is obtained 
when using a density dependence predicted by a perturbative chiral quark model for the four-quark condensate. The 
latter result points toward the possibility that the $\Lambda_c$ might be bound to a nucleus, provided,
of course, it can be produced almost at rest in the nucleus. The authors
of Ref.~\cite{Ohtani:2017wdc} still perform a similar analysis for the $\Lambda$ hyperon and find that when
the factorized four-quark condensate is employed, a very strong repulsion is obtained and thereof 
favors the unfactorized four-quark condensate, which predicts a weak attraction, in qualitative 
agreement with the mass shift of $\Lambda$ in nuclear matter as extracted from the binding energies
of hypernuclei. 

Clearly, these studies reveal that our knowledge on in-medium mass of the $\Lambda_c$ is still very 
rudimentary and suggest the need of further consideration. As already mentioned, our ability of making 
first-principles, analytical calculations of nonperturbative QCD phenomena is very limited 
and, thus, the use of models is still a valid alternative for making progress. Within such a perspective, 
in the present work we employ a widely used chiral constituent quark model~\cite{Val05,Vij05} to evaluate the
in-medium masses of charmed hadrons as well as their in-medium low-energy interactions with nucleons 
and other charmed baryons. The model provides a very good description of the low-lying spectrum of 
the light and charmed hadrons~\cite{Vac05,Val08}. The vacuum values of the parameters of 
the model, which are the masses of constituent quarks, the $\pi$ and $\sigma$ meson masses 
and their coupling constants to the light constituent quarks are therefore well constrained. 
To evaluate the temperature and baryon density dependence of those parameters we employ the 
Nambu--Jona--Lasinio (NJL) model~\cite{Nam61,Nab61}, following the strategy set up in our 
previous work in Ref.~\cite{Car16} on the in-medium properties of a $\Delta\bar{D}^*$ molecule. 

The paper is organized as follows. In Sec.~\ref{secIIB} we outline the basic ingredients of the 
chiral constituent quark model used for the study of the one- and two-baryon problems. 
In Sec.~\ref{secIII} we present and discuss the results for the in-medium masses of the charmed baryons 
$Y_c = (\Lambda_c, \Sigma_c,\Sigma^*_c)$. In Sec.~\ref{secIV} we present 
numerical results for the in-medium $\Lambda_c N$ and $\Lambda_c \Lambda_c$ interactions
in comparison to other approaches in the literature. Finally, 
in Sec.~\ref{secV} we summarize the main conclusions of our work. 

\section{Quark-quark and baryon-baryon interactions}
\label{secIIB}

In this section we define the chiral constituent quark model used in the present work~\cite{Val05}. 
The model was proposed in the early 1990s in an attempt to obtain a simultaneous description 
of the light baryon spectrum and the nucleon-nucleon interaction. It was later on generalized 
to all flavor sectors~\cite{Vij05}. In this model, hadrons are described as clusters of three 
interacting  massive (constituent) quarks. The masses of the quarks are generated by the 
dynamical breaking of the original $SU(2)_{L}\otimes SU(2)_{R}$ chiral symmetry of the QCD 
Lagrangian at a momentum scale of the order of $\Lambda_{\rm CSB} = 4\pi f_\pi \sim 1$~GeV, 
where $f_\pi$ is the pion electroweak decay constant. For momenta typically below that 
scale, when using the linear realization of chiral symmetry, light quarks interact through 
potentials generated by the exchange of pseudoscalar Goldstone 
bosons ($\pi$) and their chiral partner ($\sigma$): 
\begin{equation}
V_{\chi}(\vec{r}_{ij})\, = \, V_{\sigma}(\vec{r}_{ij}) \, + \, V_{\pi}(\vec{r}_{ij}) \, ,
\end{equation}
where
\begin{eqnarray}
V_{\sigma}(\vec{r}_{ij}) &=&
    -\dfrac{g^2_{\rm ch}}{{4 \pi}} \,
     \dfrac{\Lambda^2}{\Lambda^2 - m_{\sigma}^2}
     \, m_{\sigma} \, \left[ Y (m_{\sigma} \,
r_{ij})-
     \dfrac{\Lambda}{{m_{\sigma}}} \,
     Y (\Lambda \, r_{ij}) \right] \,, \nonumber \\
V_{\pi}(\vec{r}_{ij})&=&
     \dfrac{ g_{\rm ch}^2}{4
\pi}\dfrac{m_{\pi}^2}{12 M_i M_j}
     \dfrac{\Lambda^2}{\Lambda^2 - m_{\pi}^2}
m_{\pi}
     \Biggr\{\left[ Y(m_{\pi} \,r_{ij})
     -\dfrac{\Lambda^3}{m_{\pi}^3} Y(\Lambda
\,r_{ij})\right]
     \vec{\sigma}_i \cdot \vec{\sigma}_j 
\nonumber \\
&&   \qquad\qquad +\left[H (m_{\pi} \,r_{ij})
     -\dfrac{\Lambda^3}{m_{\pi}^3} H(\Lambda
\,r_{ij}) \right] S_{ij}
     \Biggr\}  (\vec{\tau}_i \cdot \vec{\tau}_j)
\, .
\end{eqnarray}
$g^2_{\rm ch}/4\pi$ is the chiral coupling constant, $M_i = (M_u,M_d)$ are the
masses of the constituent quarks, $\Lambda \sim \Lambda_{\rm CSB}$,  
$Y(x)$ is the standard Yukawa function defined by $Y(x)=e^{-x}/x$, 
$H(x)=(1+3/x+3/x^2)\,Y(x)$, and $S_{ij} \, = \, 3 \, ({\vec \sigma}_i \cdot
{\hat r}_{ij}) ({\vec \sigma}_j \cdot  {\hat r}_{ij})
\, - \, {\vec \sigma}_i \cdot {\vec \sigma}_j$ is
the quark tensor operator.

Perturbative QCD effects are taken into account through the one-gluon-exchange (OGE) 
potential~\cite{Ruj75}:
\begin{equation}
V_{\rm OGE}({\vec{r}}_{ij}) =
        {\frac{\alpha_s}{4}}\,{\vec{\lambda}}_{i}^{\rm
c} \cdot {\vec{\lambda}}_{j}^{\rm c}
        \Biggl[ \frac{1}{r_{ij}}
        - \dfrac{1} {4} \left(
{\frac{1}{{2\,M_{i}^{2}}}}\, +
{\frac{1}{{2\,M_{j}^{2}}}}\,
        + {\frac{2 \vec \sigma_i \cdot \vec
\sigma_j}{3 M_i M_j}} \right)\,\,
          {\frac{{e^{-r_{ij}/r_{0}}}}
{{r_{0}^{2}\,\,r_{ij}}}}
        - \dfrac{3 S_{ij}}{4 M_i M_j r_{ij}^3}
        \Biggr] \, ,
\label{OGE}
\end{equation}
where $\lambda^{c}$ are the $SU(3)$ color matrices, 
$r_0=\hat r_0/\nu$ is a flavor-dependent regularization scaling with the 
reduced mass $\nu$ of the interacting pair, and $\alpha_s$ is the
scale-dependent strong coupling constant given by~\cite{Vij05},
\begin{equation}
\alpha_s(\nu)={\frac{\alpha_0}{\rm{ln}\left[{({\nu^2+\mu^2_0})/
\gamma_0^2}\right]}},
\label{asf}
\end{equation}
where $\alpha_0=2.118$, 
$\mu_0=36.976$ MeV and $\gamma_0=0.113$ fm$^{-1}$. This equation 
gives rise to $\alpha_s\sim0.54$ for the light-quark sector,
$\alpha_s\sim0.43$ for $uc$ pairs, and
$\alpha_s\sim0.29$ for $cc$ pairs.

Finally, any model imitating QCD should incorporate
confinement. Although it is a very important term from the spectroscopic point of view,
it is negligible for the hadron-hadron interaction. Lattice calculations 
suggest a screening effect on the potential when increasing the interquark 
distance~\cite{Bal01} which is modeled here by, 
\begin{equation}
V_{\rm CON}(\vec{r}_{ij})= -a_{c}\,(1-e^{-\mu_c\,r_{ij}})\,
(\vec{\lambda^c}_{i}\cdot \vec{ \lambda^c}_{j})\,,
\end{equation}
where $a_{c}$ and $\mu_c$ are the strength and range parameters.
Once perturbative (one-gluon exchange) and nonperturbative (confinement and
dynamical chiral symmetry breaking) aspects of QCD have been incorporated, 
one ends up with a quark-quark interaction of the form,
\begin{equation} 
V_{q_iq_j}(\vec{r}_{ij})=\left\{ \begin{array}{ll} 
\left[ q_iq_j=nn \right] \Rightarrow V_{\rm CON}(\vec{r}_{ij})+V_{\rm OGE}(\vec{r}_{ij})
+V_{\chi}(\vec{r}_{ij}) &  \\ 
\left[ q_iq_j=cn/cc \right]  \Rightarrow V_{\rm CON}(\vec{r}_{ij})+V_{\rm OGE}(\vec{r}_{ij}) &
\end{array} \right.\,,
\label{pot}
\end{equation}
where $n$ stands for the light quarks $u$ and $d$.
Notice that for the particular case of heavy quarks ($c$ or $b$) chiral symmetry is
explicitly broken and therefore boson exchanges associated with the dynamical breaking
of chiral symmetry do not contribute.
The parameters of the model are the same that have been used for the
study of the one- and two-baryon systems in vacuum, and for completeness 
are quoted in Table~\ref{t1}.
\begin{table}[htb]
\caption{Quark-model parameters.}
\label{t1}
\begin{ruledtabular}
\begin{tabular}{lc|lc}
 $M_{u,d} ({\rm MeV})$    & 313    & $g_{\rm ch}^2/(4\pi)$      & 0.54  \\ 
 $m_c ({\rm MeV})$        & 1752   & $m_\sigma ({\rm fm^{-1}})$ & 3.42 \\ 
 $b ({\rm fm})$           & 0.518  & $m_\pi ({\rm fm^{-1}})$    & 0.70  \\ 
 $b_c ({\rm fm})$         & 0.6    & $\Lambda ({\rm fm^{-1}})$  & 4.2  \\ 
 $\hat r_0$ (MeV fm)      & 28.170 & $a_c$ (MeV)                & 230 \\ 
 $\mu_c$ (fm$^{-1}$)      & 0.70   &                            &  \\ 
\end{tabular}
\end{ruledtabular}
\end{table}

In order to derive the $B_n B_m\to B_k B_l$ interaction from the
basic $qq$ interaction defined above, we use a Born-Oppenheimer
approximation where the quark coordinates are integrated out keeping $R$
fixed, the resulting interaction being a function of the two-baryon relative 
distance. A thorough discussion of the model can be found elsewhere~\cite{Val05,Vij05}. 

%
%
\section{In-medium charmed baryon masses}
\label{secIII}

Medium effects on the baryon masses and baryon-baryon interactions are incorporated 
within a quasiparticle picture of the nuclear many-body problem, in which
the parameters of the underlying quark model carry the medium effects~\cite{Car16}. 
This is similar to the calculations of $Y_c$ masses in the quark-meson coupling model~\cite{Tsushima:2002ua,Tsu04}, 
in which meson mean fields sourced by quark scalar and vector in-medium densities couple to current quarks within 
the baryons, and also in QCD sum rules~\cite{Wan11,Wan12,{Azizi:2018dtb}, {Azi17},{Ohtani:2017wdc}}, in which medium 
effects are carried by (quark, gluon and mixed quark-gluon) condensates. In the chiral quark model that we employ here, 
the temperature, $T$, and baryon density, $\rho_B$, dependence of model parameters are those predicted by the NJL model.
This choice is motivated by the fact that the bosonized version of the NJL model with $\sigma$ and $\pi$ 
mesons~\cite{Eguchi:1976iz,Kle92} leads to the same Yukawa quark-meson couplings as those in the 
chiral constituent quark model discussed in Sec.~\ref{secIIB}. In addition, it gives very 
simple expressions for the masses and quark-meson couplings in vacuum and also for nonzero
$T$ and $\mu$. Moreover, for sufficiently low 
values of $T$ and $\rho_B$, the NJL model reproduces~\cite{Vog91,Kle92,Hat94,Bub05} the 
model-independent predictions derived in the context of chiral
perturbation theory for the in-medium quark condensate
$\tave{\bar q q}$~\cite{Ger89,Dru91,Cohen:1994wm}:
\begin{equation}
\frac{\tave{\bar q q}}{\ave{\bar q q}} \simeq 1 - \frac{T^2}{8f^2_\pi} 
- \frac{1}{3} \, \frac{\rho_B}{\rho_0} \, ,
\label{cond-Trho}
\end{equation}
where $\langle\bar q q \rangle$ is the vacuum light quark condensate, $f_\pi$ the vacuum
pion electroweak decay constant, and $\rho_0$ the baryon saturation density of nuclear 
matter. For $\rho_B = 0$, this prediction would be valid for $T \lesssim 0.1$~GeV~\cite{Ger89}. In a treatment that
includes thermal excitations of the pions, the NJL model reproduces Eq.~(\ref{cond-Trho}) 
very well up to $T \simeq 0.1$~GeV~\cite{Bub05}. When the model is solved in the Hartree 
approximation, the one used in the present work, it gives a value for 
$\tave{\bar q q}/\ave{\bar q q}$ that is 10\% larger than given by Eq.~(\ref{cond-Trho}).
In the particular case of $\rho_B/\rho_0 = 0.5$, which corresponds to $\mu =0.19$~GeV, 
it is 25\% larger. While the pion mass is 
protected by chiral symmetry, i.e., its mass does not change while chiral symmetry is not restored, the 
masses of the constituent quarks and of the $\sigma$ meson are to a good approximation proportional 
to the quark condensate. Despite the existence of these model independent results, a model is still 
required mainly because further input is needed, namely the $T$ and $\rho_B$ dependence of the quark-meson 
coupling constants. The relevant equations determining the $T$ and $\rho_B$ dependence of the masses and 
couplings are well known since long time~\cite{Vog91,Kle92,Hat94,Bub05}. The particular implementation 
of the model is the one used in our previous work in Ref.~\cite{Car16}. We refer the reader to that 
reference for details and also discussions on the limitations of the calculations.   

\begin{figure*}[t]
\vspace*{-0.25cm}
\hspace*{-0.5cm}\resizebox{8.5cm}{13cm}{\includegraphics{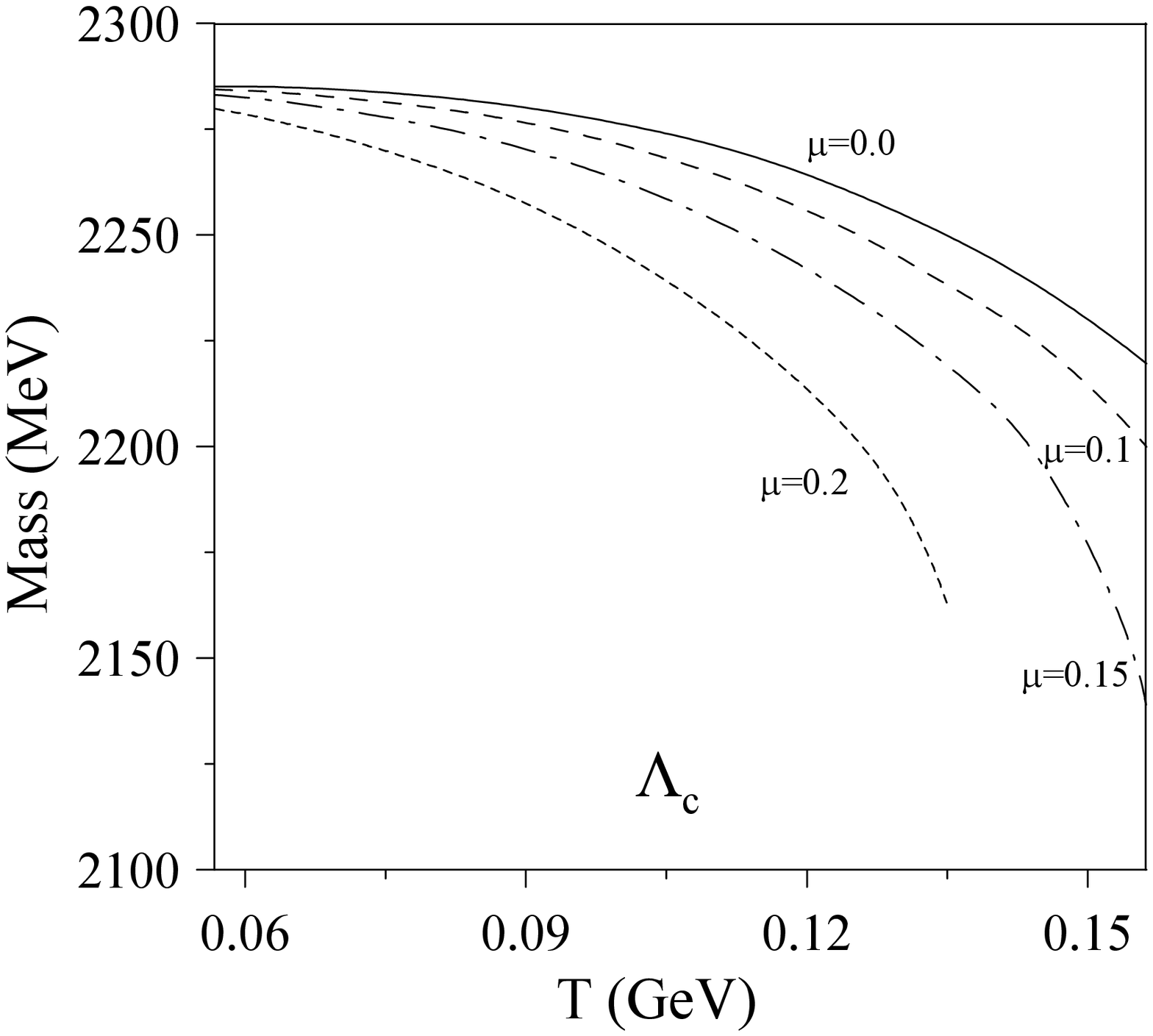}}
\hspace*{-0.5cm}\resizebox{8.5cm}{13cm}{\includegraphics{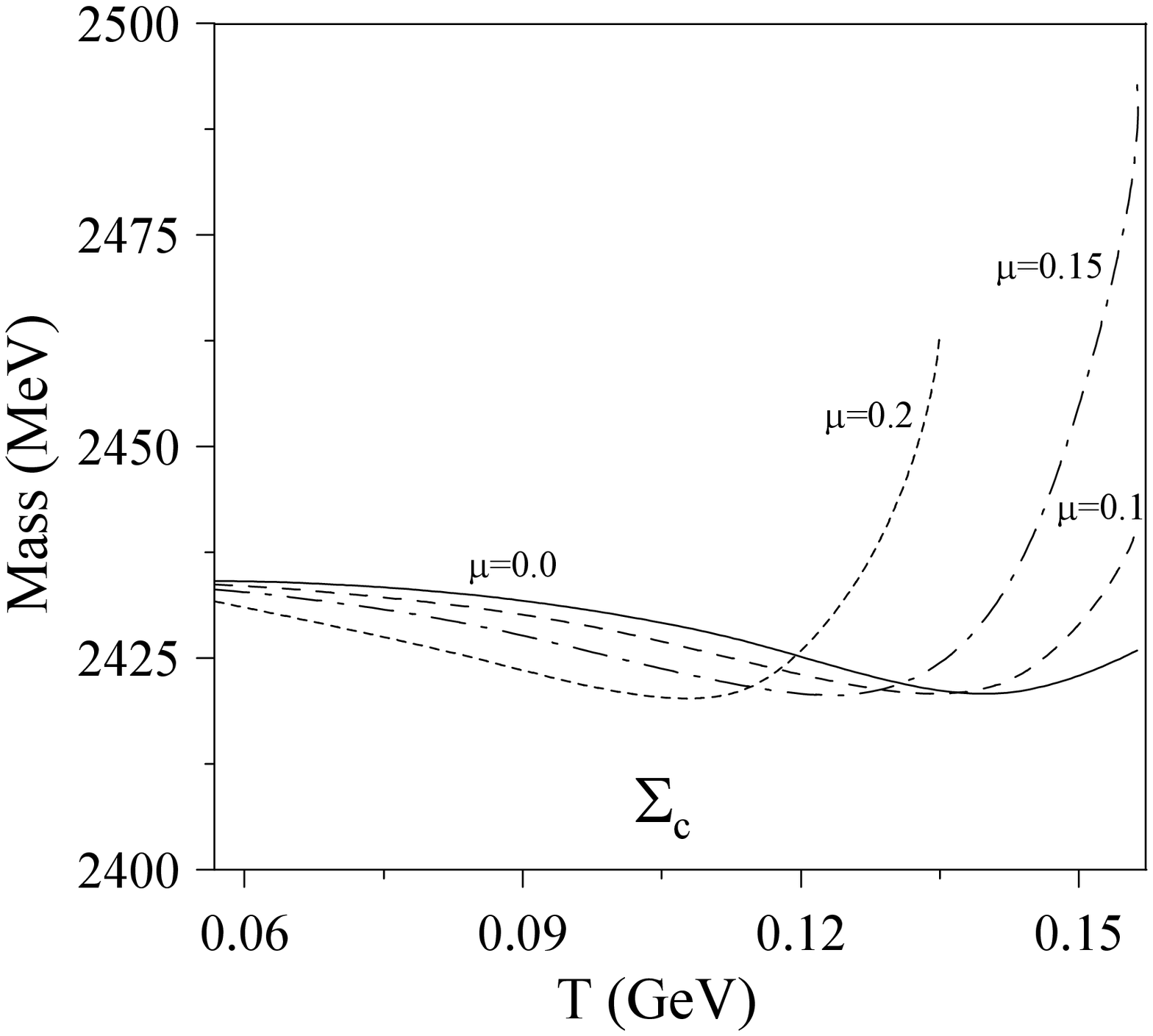}}\vspace*{-5cm}
\hspace*{-0.5cm}\resizebox{8.5cm}{13.cm}{\includegraphics{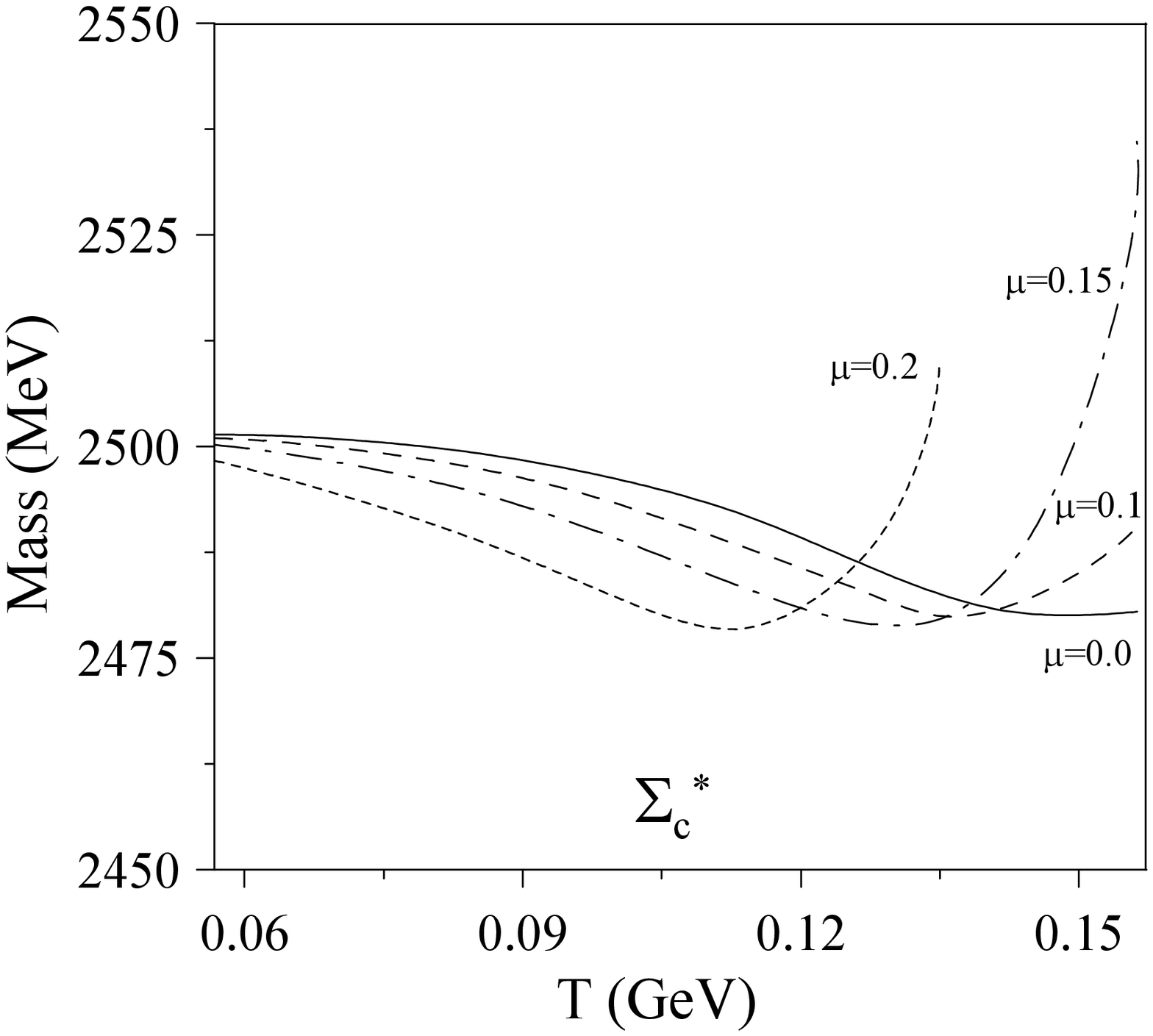}}
\vspace*{-5.5cm}
\caption{Masses of $\Lambda_c$, $\Sigma_c$ and $\Sigma_c^*$ baryons as a function of 
the temperature for the different values of the baryon chemical potential $\mu$ (in GeV). }
\label{fig2}
\end{figure*}

The in-medium dependence of the charmed baryon masses are readily obtained by solving
the bound-state problem of three constituent quarks as detailed in Refs.~\cite{Vac05,Val08}
with the temperature, $T$, and baryon chemical potential, $\mu$, dependence of the quark and 
meson masses and the quark-meson couplings as obtained in Ref.~\cite{Car16}.
In Fig.~\ref{fig2} we depict the variation of the masses of 
$\Lambda_c$, $\Sigma_c$, and $\Sigma_c^*$ baryons as a function of $T$ for different values of 
the baryon chemical potential $\mu$. 
As one can see, we find that the mass of the $\Lambda_c$ decreases monotonically with both $T$ and 
$\mu$, in qualitative agreement with the results of Refs.~\cite{Tsu04,{Azi17}} and those of 
Ref.~\cite{Ohtani:2017wdc} when using the unfactorized four-quark condensate with a density dependence 
taken from a perturbative chiral quark model. In more quantitative terms, our model 
predicts a mass decrease of $\Delta M_{\Lambda_c} = 72$~MeV at $\rho_B/\rho_0 = 1$, which 
at the highest value of $\mu$ used, $\mu=0.2$~GeV, corresponds to $T \simeq 0.12$~GeV. At this
values, ${\tave{\bar q q}}/{\ave{\bar q q}} \simeq 0.7$. This mass decrease is smaller
than the one obtained by the QMC model~\cite{Tsu04}, $\Delta M_{\Lambda_c} \simeq 122$~MeV, 
and much larger than those obtained in the QCD sum rules calculation of Refs.~\cite{Azi17,Ohtani:2017wdc}, 
which give $\Delta M_{\Lambda_c} \simeq 10$~MeV. Although in the QCD sum rules calculations, 
${\tave{\bar q q}}/{\ave{\bar q q}}$ at $\rho_B/\rho_0 = 1$ is essentially equal to the one in
our calculation, the results show that there are differences in the way chiral symmetry restoration
works in the different models. While in the present calculation the interpretation for the decrease
in the mass of $\Lambda_c$ can be made (see the discussion in the next paragraph) in terms of 
the interplay between increased kinetic energy and the spin-spin interactions due to gluon and 
pion exchanges, in the QCD sum rules such an assessment is more difficult, as discussed e.g. in 
Ref.~\cite{Ohtani:2017wdc}. In the case of the QMC model, a role similar to the quark condensate is played 
by a scalar nuclear mean field. The interplay between increased kinetic terms and attractive 
zero-point and c.m. energies is clearly exposed in the hadron mass formulae~\cite{Tsu04}. On the other 
hand, the QCD sum rule conclusions of Refs.~\cite{Wan11,Wan12} and those of Ref.~\cite{Ohtani:2017wdc} based on the 
factorized four-quark condensate disagree with our predictions. At this point, it needs to be said 
that Ref.~\cite{Ohtani:2017wdc} notes how the density dependence of the factorized four-quark condensate is 
too strong to explain the observed binding of $\Lambda$ in nuclei, a feature which advocates in favor
of the unfactorized weak density dependence of the perturbartive chiral quark model. On the other 
hand, the masses of $\Sigma_c$ and $\Sigma^*_c$ do not decrease monotonically with~$T$, they start 
decreasing for small values of $T$ but then turn over and become larger than their vacuum values for 
higher values of~$T$. The turnover temperature decreases as $\mu$ increases.

The behavior of the in-medium masses of charmed baryons can be easily understood in 
terms of a simple group theory analysis~\cite{Val14}.
In an approximation in which the heavy-quark masses are taken $M_Q \to \infty$, 
the angular momentum of the light degrees of freedom is a good 
quantum number. Thus, heavy-quark baryons belong either to the flavor $SU(3)$ antisymmetric 
$\mathbf{\bar{3}_F}$ representation, or to the symmetric $\mathbf{6_F}$ representation. 
The spin of the light diquark is 0 for $\mathbf{\bar{3}_F}$, while it is 1
for $\mathbf{6_F}$. Thus, while the spin of the ground
state baryons is $1/2$ for the $\mathbf{\bar{3}_F}$ representation, which
contains among others the $\Lambda_c$ baryon, it can be
both $1/2$ or $3/2$ for the $\mathbf{6_F}$ representation, which 
contains among others the $\Sigma_c$ and the $\Sigma_c^*$, respectively. 
Therefore, heavy hadrons would form doublets, in that $\Sigma_Q$ and $\Sigma_Q^*$  
would be mass-degenerate in the limit $M_Q \rightarrow \infty$ and, away from
this limit, there is a mass splitting due to the spin-spin interaction at 
order $1/M_Q$. The mass difference between states belonging to the flavor
$\mathbf{\bar{3}_F}$ and $\mathbf{6_F}$ representations
tends to a constant when the heavy quark mass $M_Q \to \infty$,
due to the dynamics of the light diquark subsystem, so that:
\begin{eqnarray}
{\rm M}[\Sigma_c^*] - {\rm M}[\Sigma_c] &\Rightarrow & 
\Delta{\rm M}\left( [\mathbf{6_F}] - [\mathbf{6_F}]\right) \equiv V_{\text{light-charm}} \nonumber \\
{\rm M}[\Sigma_c^*] - {\rm M}[\Lambda_c] &\Rightarrow & 
\Delta{\rm M}\left( [\mathbf{6_F}] - [\mathbf{\bar{3}_F}]\right) \equiv V_{\text{light-light}} \, .
\label{Sp-Sp}
\end{eqnarray}
Let us note that in $\Lambda_c$ there is an attractive
{\it ud} diquark (``good" diquark) with color $\mathbf{\bar{3}}$, spin
0 and isospin 0, whereas in $\Sigma_c$ and
$\Sigma_c^*$ there is a repulsive
{\it ud} diquark (``bad" diquark) with color $\mathbf{\bar{3}}$, but 
spin 1 and isospin 1. This is why $\Sigma_c$ and $\Sigma_c^*$ baryons 
follow a similar behavior with temperature and density whereas the 
$\Lambda_c$ has a completely different behavior, its mass diminishing due
to the attractive character of the one-gluon exchange for a spin zero diquark,
effect that is increased when the mass of the quark diminishes. 
As can be seen in Eq.~(\ref{Sp-Sp}), the mass difference between the 
members of the $\mathbf{\bar{3}_F}$ and $\mathbf{6_F}$ comes determined
by the dynamics of the two-light quarks. Being the spin-isospin pairs in
a symmetric state for both configurations it is the switch of the
spin-color pairs symmetry, symmetric for the flavor $\mathbf{\bar{3}_F}$
representation and antisymmetric for the $\mathbf{6_F}$, the responsible
for the nonmonotonic behavior of the $\Sigma_c$ and
$\Sigma_c^*$ masses. While in the $\Lambda_c$ all contributions (except for the kinetic
energy) are attractive, in the members of the $\mathbf{6_F}$ representation
the spin-color interaction becomes repulsive. For small variations of the 
mass of the light quarks there is a compensation between the attractive
character of the pseudo-Goldstone boson exchange interaction and the repulsive
OGE and kinetic energy contributions. For larger temperatures, for
a given baryon chemical potential, the repulsive character
of the spin-color interaction, depending on the regularization
of a $\delta$-function through the reduced mass of the interacting quarks 
[see Eq.~(\ref{OGE})], together with the increase of the
repulsive kinetic energy, dominates the attractive contributions.
For the $\Lambda_c$ case, the spin-color interaction is also attractive,
increasing in this way the slope of the decrease of the $\Lambda_c$
mass when the $\Sigma_c$ and $\Sigma_c^*$ masses present the turnover.

These results show that charmed baryons offer an ideal laboratory for 
learning on temperature and density effects on the phenomenon of dynamical chiral
symmetry breaking. The different two-quark subsystems, heavy-light and light-light,
are clearly disentangled by the way they react to changes in the quark masses,
which in the case of baryons affects primarily the spin-spin interaction~\cite{Car16}. 

%
%
\section{In-medium binding of charmed baryons}
\label{secIV}

In this section we investigate how the interaction of the $\Lambda_c$ baryon with
nucleons and with other charmed baryons is modified in a medium at finite $T$ and $\mu$. 
We note that our calculation is particularly applicable for a medium similar to the one formed in a high-energy 
heavy-ion collision, in which quarks coalesce to form weakly bound hadron molecules~\cite{Cho}. 
At finite temperature and/or baryon chemical potential there are several competing
effects. On one hand, the interactions grow due to the decrease 
in $m_\sigma$ at finite $T$ and $\mu$ providing a stronger interaction. Besides, 
thresholds are modified due to the changes in the masses of the charmed baryons,
shown in Fig.~\ref{fig2}, and the nucleon, see Fig. 2(b) of Ref.~\cite{Car16}. 
Finally, coupled-channel effects through the $\Lambda_c \leftrightarrow \Sigma_c$ 
conversion are less important because,
as seen in the previous section, the mass difference increases both with the
temperature and the baryon chemical potential. In the following, we investigate the $T$ and $\mu$ dependence
of the $\Lambda_c N$ and $\Lambda_c \Lambda_c$ interactions taking into account all those effects.
We note that in the present model these two-baryon systems are not bound in vacuum~\cite{Gar15,Car15}.

To study the possible existence of two-baryon bound states, we solve the 
Lippmann-Schwinger equation for negative energies as has been detailed in Ref.~\cite{Car16} 
and examine the Fredholm determinant $D_F(E)$ at zero energy~\cite{Gar87}. 
For noninteracting systems $D_F(0)=1$, for an attractive two-baryon 
interaction $0<D_F(0)<1$, and when there exits a bound state $D_F(0)<0$. 
\begin{figure}[t]
\hspace*{-1cm}\mbox{\epsfxsize=100mm\epsffile{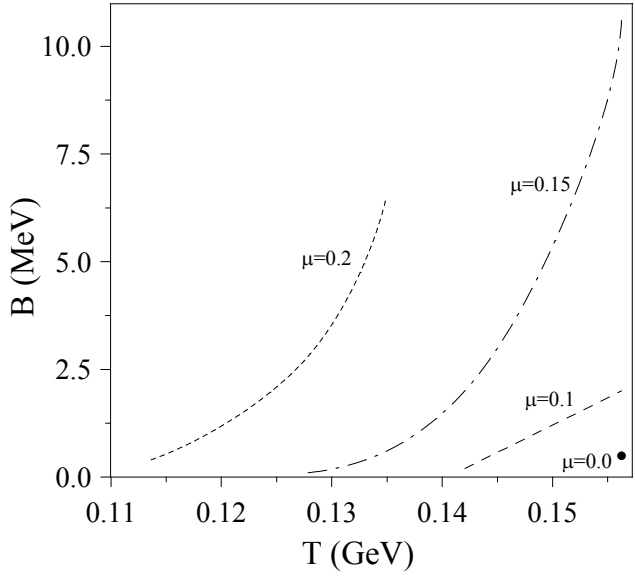}}
\vspace*{-5.5cm}
\caption{Binding energy of the $(I)J^P=(1/2)1^+$ $N\Lambda_c$ state, 
as a function of the temperature $T$ for different values of the baryon 
chemical potential $\mu$ (in GeV).}
\label{fig3}
\end{figure}
Making use of the in-medium baryon masses and baryon-baryon interactions obtained with the
chiral constituent quark model with $T-$ and $\mu-$dependent parameters, we have analyzed 
the lowest $N\Lambda_c $ states, $(I)J^P=(1/2)0^+$ and $(I)J^P=(1/2)1^+$. 
The $(I)J^P=(1/2)0^+$ channel is always repulsive. 
We show in Fig.~\ref{fig3} the binding energy of the $(I)J^P=(1/2)1^+$ $N\Lambda_c$ state 
as a function of $T$ for different values of~$\mu$. In all cases $B=0$
corresponds to the mass of the corresponding threshold, i.e., 
$M_{N}(T,\mu)+M_{\Lambda_c}(T,\mu)$. It can be seen how the $\Lambda_c$
starts to be bound as $T$ and $\mu$ increase. In fact, for $\mu=0$ there is only a tiny 
binding for the largest temperature considered. The values of the binding energies are small despite the 
large decrease in the masses of the nucleon and the $\Lambda_c$.
Note that the largest binding obtained is of the order of 10 MeV, a value comparable to the 
binding energies of charmed hypernuclei in Refs.~\cite{{Tsushima:2002ua},Tsu04,Gar15} 
and also with the 20 MeV mass shift obtained in Ref.~\cite{Ohtani:2017wdc} at the normal
nuclear matter density when using the unfactorized four-quark
condensate. We can identify several competing effects that add or cancel to arrive 
to this final value. First of all, we note that the $\Lambda_c \leftrightarrow \Sigma_c$ 
conversion is less important than in the similar system in strange sector, mainly due 
to the larger vacuum mass difference, namely 168~MeV as compared to 73 MeV in the strange sector.
Besides, it comes reduced as compared to the strange sector due to 
the absence of strange meson exchanges~\cite{Ban83}, giving rise to a 
smaller $N\Lambda_c \leftrightarrow N\Sigma_c$ transition potential.
Finally, the $\Lambda_c \leftrightarrow \Sigma_c$ conversion comes also suppressed
when increasing $T$ and/or $\mu$, due to a larger mass difference, 
as seen in Fig.~\ref{fig2}. However, the increase of the interacting potential due to the 
decrease of $m_\sigma$ is enough to give binding despite the smaller threshold mass, 
increasing the kinetic energy contribution.

\begin{figure}[t]
\hspace*{-1cm}\mbox{\epsfxsize=100mm\epsffile{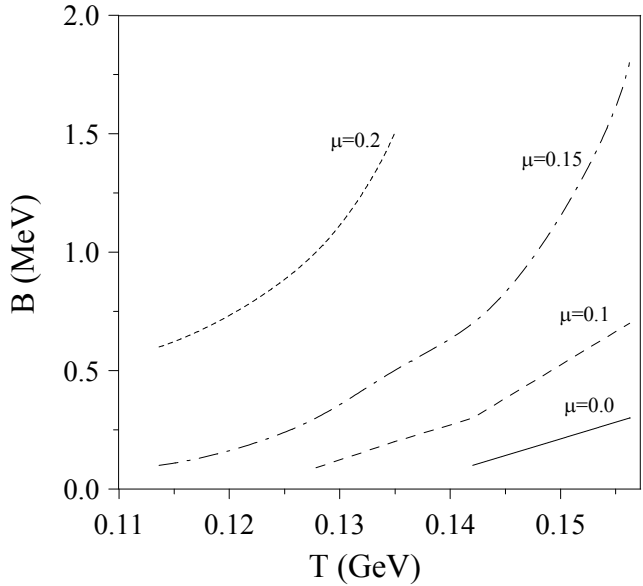}}
\vspace*{-5.5cm}
\caption{Binding energy of the $(I)J^P=(0)0^+$ $\Lambda_c\Lambda_c$ system, 
as a function of the temperature $T$ for different values of the 
baryon chemical potential $\mu$ (in GeV).}
\label{fig4}
\end{figure}

We have performed the same analysis for $\Lambda_c\Lambda_c$ system.
We show in Fig.~\ref{fig4} the results for the lowest channel, $(I)J^P=(0)0^+$.
This system was studied in vacuum in Ref.~\cite{Car15}, concluding the 
nonexistence of a charmed $H$-like dibaryon, although it may
appear as a resonance above the $\Lambda_c\Lambda_c$ threshold.
It is important to note that the $\Lambda_c\Lambda_c$ system is decoupled from the
closest two-baryon threshold, the $N\Xi_{cc}$ state, that in the case of the
strange $H$ dibaryon becomes relevant for its possible bound or resonant character~\cite{Ino12}. 
The binding of the $(I)J^P=(0)0^+$ state would then require a stronger attraction
in the diagonal channels or a stronger coupling to the heavier $\Sigma_c\Sigma_c$ state. 
However, the mass difference between the two coupled channels in 
the $(I)J^P=(0)0^+$ partial wave, namely the $\Lambda_c\Lambda_c$ and $\Sigma_c\Sigma_c$, are 
much larger than in the counterpart strange sector, and increase both with 
$T$ and $\mu$, making the coupled-channel effect less important.
Let us note that in the strange sector one has $M(N\Xi) - M(\Lambda\Lambda)=$ 25 MeV and 
$M(\Sigma\Sigma) - M(\Lambda\Lambda)=$ 154 MeV. In the charm sector,
the closest channel coupling to $\Lambda_c\Lambda_c$ in the $(I)J^P=(0)0^+$ state is
$\Sigma_c\Sigma_c$, which in vacuum is 338~MeV above. 

Thus, when increasing the temperature and/or the chemical potential, the $\Lambda_c\Lambda_c$ 
and $\Sigma_c\Sigma_c$ thresholds are separated and, therefore, not much further binding can
be expected from coupled-channel effects. Furthermore, the decrease of $\Lambda_c$ mass  
increases the repulsion due to an increase in the kinetic energy and these two effects 
can only be compensated by the increase of the interacting potential. The net effect is that 
the in-medium $\Lambda_c \Lambda_c$ binding is much smaller than the in-medium $\Lambda_c N$
binding.

%
%
\section{Summary}
\label{secV}

In brief, we have studied the effect of temperature $T$ and baryon chemical potential
$\mu$ on the masses of the $\Lambda_c$, $\Sigma_c$ and $\Sigma^*_c$ charmed baryons 
and on the $\Lambda_c N$ and $\Lambda_c \Lambda_c$ interactions. We have used a chiral 
constituent quark model, in which the parameters are taken to be 
$T-$ and $\mu-$dependent as predicted by the Nambu-Jona-Lasinio model. 
We have found that while the mass of the $\Lambda_c$ baryon decreases monotonically
in medium, the masses of the $\Sigma_c$ and $\Sigma^*_c$ have a nonmonotonic behavior.  
We have shown that the behavior of the in-medium masses of 
those baryons can be understood in terms of a simple group theory analysis, which 
allows us to disentangle the dynamics of the different heavy-light and light-light
two-quark subsystems composing the baryons. Thus, these systems offer a unique
laboratory for learning on temperature and density effects on the phenomenon of 
dynamical chiral symmetry breaking. We have compared our results with others in 
the literature using different models, relativistic mean field models for nuclear 
matter and QCD sum rules. 

Regarding the $\Lambda_c N$ and $\Lambda_c \Lambda_c$ in-medium interactions,
we found that there is a delicate balance involving a stronger interaction due
to the decrease of the mass of $\sigma$, a suppression of coupled-channel effects 
and lighter thresholds, leading to an overall effective attraction. We have found 
an in-medium $\Lambda_c N$ binding energy of the order of 10 MeV, in qualitative agreement
with a calculation using QCD sum rules in which the density dependence 
of the unfactorized four-quark condensate is estimated from a perturbative chiral 
quark model. Our result clearly points to the possibility that the $\Lambda_c$ can 
be bound in a sufficiently large nucleus, as a binding energy of 10~MeV is comparable
to the binding energies found in calculations of $\Lambda_c$ hypernuclei using relativistic 
mean field. For the $\Lambda_c \Lambda_c$ system, we found a shallow bound state with a binding 
energy of the order of 2 MeV. Such systems can in principle be formed through coalescence in the
environment of a heavy-ion collision. The general conclusion is that the $\Lambda_c N$ 
and $\Lambda_c \Lambda_c$ systems that are not bound in vacuum, could become bound in
a medium at finite temperature and finite baryon density. Our findings are relevant 
for ongoing heavy-ion experiments at the Relativistic Heavy-Ion Collider
(RHIC) and the LHC, and for the planned experiments at FAIR and J-PARC.

As already mentioned, charmed hadrons can be produced and their interactions realistically 
measured in high-energy heavy-ion collisions. In addition, the future research programs at different 
facilities like FAIR and J-PARC are expected to improve our knowledge on the in-medium 
hadron-hadron interactions involving heavy flavors. While the scarce experimental information 
leaves room for some degree of speculation in the study of processes involving charmed hadrons, 
the situation can be ameliorated with the use of well constrained models based as much as possible 
on symmetry principles and analogies with other similar processes. The present detailed theoretical 
investigation of the behavior of the in-medium masses of charmed hadrons and their interactions
is based on well established models. It is hoped that our work is of help toward raising 
the awareness of experimentalists that it is worthwhile to investigate few-baryon systems
involving heavy-flavor hadrons, specifically because for some quantum numbers such states 
could form interesting bound states. 

%
\section{acknowledgments}
This work has been partially funded
by Ministerio de Econom\'\i a, Industria y Competitividad 
and EU FEDER under Contract No. FPA2016-77177
and by a bilateral agreement Universidad de 
Salamanca - Funda\c{c}\~ao de Amparo \`a Pesquisa do Estado de S\~ao Paulo - FAPESP Grant 
No. 2017/50269-7. Partial financial support is also acknowledged from Conselho 
Nacional de Desenvolvimento Cient\'{\i}fico e Tecnol\'ogico - CNPq, 
Grants No. 168445/2017-4 (C.E.F.), 305894/2009-9 (G.K.), 464898/2014-5(G.K) (INCT F\'{\i}sica 
Nuclear e Apli\-ca\-\c{c}\~oes), and Funda\c{c}\~ao de Amparo \`a 
Pesquisa do Estado de S\~ao Paulo - FAPESP, Grant No. 2013/01907-0 (G.K.). 
%

%

\end{document}